\documentstyle[sprocl]{article}

\def\Journal#1#2#3#4{{#1} {\bf #2}, #3 (#4)}

\def\PLB{{\em Phys. Lett.}  B}
\def\PRL{\em Phys. Rev. Lett.}
\def\PRD{{\em Phys. Rev.} D}

\def\be{\begin{equation}}
\def\ee{\end{equation}}
\def\bea{\begin{eqnarray}}
\def\eea{\end{eqnarray}}
\begin{document}

\title{Do internal symmetries get restored in hot and dense SUSY?}

\author{BORUT BAJC}

\address{International Centre for Theoretical Physics, 
P.O.B. 586, 34100 Trieste, Italy}

\address{Institut J. Stefan, Jamova 39, P.O. Box 3000, 1001 Ljubljana,
Slovenia}


\maketitle\abstracts{I offer some computational details and 
useful and concise formulae to calculate the effective potential 
for a general abelian supersymmetric model at high temperature and 
density. It will be shown that such cases are very good candidates for 
symmetry nonrestoration at high temperature, providing large densities 
are present.}
  
\section{Introduction}
%\subsection{Producing the Hard Copy}\label{subsec:prod}

This talk is complementary to the one given by Goran 
Senjanovi\' c~\cite{s97} at the same conference. 
The introduction, motivation, 
physical applications and main references are already given there, 
so I will try to give some technical details.

\section{The general Abelian example}

Let me consider a general supersymmetric model with $N$ chiral 
superfields, $M$ global $U(1)$ symmetries and, for simplicity's sake, 
one single gauge $U(1)$ interaction. For each continuous symmetry 
a chemical potential corresponding to the $0$-component of the 
conserved current ($=$ charge density) is introduced:

\begin{equation}
\delta{\cal
L}^{(1)}=\mu_a\left[g^a\lambda\bar\lambda+f_i^a\psi_i\bar\psi_i
+ib_i^a[\phi_i(D_0\phi_i)^*-\phi_i^*(D_0\phi_i)]\right]\;,
\label{dl1}
\end{equation}

\noindent
where $b_i^a,f_i^a$ and $g^a$ are the $a^{th}$ charges 
of the boson $\phi_i$, fermion $\psi_i$ and gaugino $\lambda$. 
The summation over repeating indices 
($a=1,...,M+1$; $i=1,...,N$) is always assumed. 
Eq.~\ref{dl1} generates~\cite{hw82,bbd91} for each boson 
the term ($q_i^a=b_i^a$ here)

\begin{equation}
\delta{\cal L}_i^{(2)}=\mu_i^2|\phi_i|^2\;,\;\,
\mu_i=q_i^a\mu_a\;.
\label{dl2}
\end{equation}

We are interested in the effective potential at temperatures T 
much bigger than any mass in the superpotential, and, 
to simplify the formulae, for chemical potentials not 
much bigger than the temperature. 
In this case, the only new relevant terms given by a nonzero 
charge are $-a\mu_i^2T^2/12$ 
for each field, where $a=1$ for fermions and $2$ for bosons.
To see it, assume for a moment that the chemical potential $\mu$ 
corresponding to a field is an independent but nonpropagating 
field itself. One then calculate the one loop high T correction 
to the $\mu^2$ term in the effective potential: two external 
"fields" $\mu$ are connected by a loop of the corresponding 
field via the "interaction" terms in Eqs.~\ref{dl1}-\ref{dl2}. 

The leading order effective potential at high temperature and density is

\begin{equation}
V_{eff}(\phi_i,T,n_i)=V_{eff}(\phi_i,T,0)-{1\over 2}
\mu^a{\cal M}_{ab}\mu^b +\mu^a n_a\;,
\label{veffmu}
\end{equation}

\noindent
where the $(M+1)\times(M+1)$ symmetric matrix ${\cal M}$ is defined as

\begin{equation}
\label{m}
{\cal M}^{ab}(\phi,T)={T^2\over 6}(f_i^af_i^b+
2b_i^ab_i^b+g^ag^b)+
2b_i^ab_i^b|\phi_i|^2\;.
\end{equation}

\noindent
The effective potential does not depend on the 
chemical potentials, so

\begin{equation}
{\partial V_{eff}\over\partial\mu^a}=0\;,\;\;
\mu^a=({\cal M}^{-1})^{ab}n_b\;,
\label{mu}
\end{equation}

%The solution for the chemical potentials immediately gives

\begin{equation}
V_{eff}(\phi_i,T,n_i)=V_{eff}(\phi_i,T,0)+{1\over 2}
n_a({\cal M}^{-1})^{ab}n_b \;.
\label{veffn}
\end{equation}

There are some points to be stressed:

\noindent
1) Nonsupersymmetric models have a very similar matrix as 
the one defined in Eq.~\ref{m}, 
except that the number of complex bosons $\phi_i$ is not
necessarily equal to the number of Weyl fermions $\psi_i$ and 
there are no gauginos $\lambda$.

\noindent
2) A chemical potential $\mu^a$ can be nonzero even if the 
corresponding charge density $n^a$ is zero, Eq.~\ref{mu}. 
So, a large charge density 
can give a nontrivial vev to fields which 
are blind to this charge. For example, one can spontaneously 
break the electromagnetic $U(1)$, even if the universe is 
electrically neutral.

\noindent
3) One could work with a nonzero vev $A_0$ instead of the local 
$U(1)$ chemical potential $\mu^{M+1}$. The equations are 
completely the same; one must just 
equate $gA_0=\mu^{M+1}$, while the charges of the boson field $\phi_i$ 
and fermion field $\psi_i$ are equal to the ``electric charge'', 
$b_i^{M+1}=f_i^{M+1}=q_i^{gauge}$. 

\noindent
4) The first part of Eq.~\ref{veffn}, i.e. $V_{eff}(\phi_i,T,0)$ 
minimizes for the symmetric vacuum, $\phi_i=0$. This is a consequence 
of the constraints of supersymmetry, the result being valid even 
for nonrenormalizable superpotentials~\cite{bs97}. Ordinary models 
behave differently and allow also nontrivial minima.

\noindent
5) Due to the previous remark, it is then the second part of 
Eq.~\ref{veffn} which decides whether we get symmetry nonrestoration 
or not. Typically one has ${\cal M}\approx T^2+|\phi|^2$, so 
that its inverse minimizes for an infinite vev. 
For small charge densities the first term in Eq.~\ref{veffn} 
dominates, while for large densities both terms are important, 
leading to symmetry nonrestoration. The critical value 
for this to happen is clearly proportional to $T^3$, 
the temperature being the only remaining scale. 
Notice that any conserved charge density in the early universe is 
proportional to $T^3$, so the ratio between the charge density in 
the universe and the critical density is a constant number 
as long as possible symmetry breaking terms at low energy 
are not important.

\section{Conclusions}

I have given some general formulae and expressions, which can be 
useful in studying the effective potential at high temperature 
and density. In all the examples checked, they lead to symmetry 
nonrestoration for sufficiently large charge densities, the 
simplest example being the one given in Ref.~\cite{rs97}. 

The above equations can be generalized for the nonabelian 
gauge symmetries. One must introduce only the chemical potentials 
corresponding to the diagonal generators~\cite{hw82}. 
Nonabelian generators are traceless, so the associated chemical 
potentials can become nonzero only after the appearence of some 
nontrivial vev. In order to calculate the first 
critical charge density (different fields get usually nonzero 
vevs at different stages), one can then ignore the chemical 
potentials corresponding to the nonabelian symmetry.

Of course, one is especially interested in realistic models, like 
for example the MSSM or SUSY GUTs. Typically, at sufficiently large 
charge densities, all the complex fields get a nonzero vev, so 
the symmetry seems to be completely broken in this case. Further studies 
need to be done, however, since the analyses of the effective potential 
become very involved once many different fields are introduced, as 
is the case with the MSSM. In any case, the answer to the 
title seems to be simply no, given a large enough density. 

\section*{Acknowledgments}
It is a pleasure to thank the organizers of COSMO97 for the 
excellent conference and for having arranged nice weather on 
the free afternoon. This work was done in collaboration with 
Antonio Riotto and Goran Senjanovi\' c, but I also enjoyed many 
useful discussions with Denis Comelli, Gia Dvali and Alejandra Melfo. 
This work was partly supported by the British Royal Society and the 
Ministry of Science and Technology of the Republic of Slovenia. I 
also thank ICTP Trieste for their hospitality, where part of this 
work was done.

\end{document}